\documentclass[runningheads]{llncs}
\usepackage{amsmath}
\usepackage{amssymb}
\usepackage{graphicx}
\usepackage{pifont}
\newcommand{\cmark}{\ding{51}}%
\newcommand{\xmark}{\ding{55}}%
\usepackage[hidelinks]{hyperref}
\usepackage{color}

\newcommand{\R}{\mathbb{R}}
\begin{document}
\title{TabMixer: Noninvasive Estimation of the Mean Pulmonary Artery Pressure via Imaging and Tabular Data Mixing}
\titlerunning{TabMixer: Imaging and Tabular Data Mixing}
%
\author{Michal K. Grzeszczyk\inst{1} \and Przemysław Korzeniowski\inst{1} \and Samer Alabed \inst{2} \and Andrew J. Swift\inst{2} \and Tomasz Trzciński \inst{3,4,5} \and Arkadiusz Sitek\inst{6}}


\institute{Sano Centre for Computational Medicine, Cracow, Poland \\
\email{m.grzeszczyk@sanoscience.org} \and
University of Sheffield, Sheffield, England \and
Warsaw University of Technology, Warsaw, Poland \and
IDEAS NCBR, Warsaw, Poland \and
Tooploox, Wroclaw, Poland \and
Massachusetts General Hospital, Harvard Medical School, Boston, MA, USA \\
}
%
\authorrunning{M.K. Grzeszczyk et al.}
%
%
\maketitle              
\begin{abstract}
Right Heart Catheterization is a gold standard procedure for diagnosing Pulmonary Hypertension by measuring mean Pulmonary Artery Pressure (mPAP). It is invasive, costly, time-consuming and carries risks. In this paper, for the first time, we explore the estimation of mPAP from videos of noninvasive Cardiac Magnetic Resonance Imaging. To enhance the predictive capabilities of Deep Learning models used for this task, we introduce an additional modality in the form of demographic features and clinical measurements. Inspired by all-Multilayer Perceptron architectures, we present TabMixer, a novel module enabling the integration of imaging and tabular data through spatial, temporal and channel mixing. Specifically, we present the first approach that utilizes Multilayer Perceptrons to interchange tabular information with imaging features in vision models. We test TabMixer for mPAP estimation and show that it enhances the performance of Convolutional Neural Networks, 3D-MLP and Vision Transformers while being competitive with previous modules for imaging and tabular data. Our approach has the potential to improve clinical processes involving both modalities, particularly in noninvasive mPAP estimation, thus, significantly enhancing the quality of life for individuals affected by Pulmonary Hypertension. We provide a source code for using TabMixer at \url{https://github.com/SanoScience/TabMixer}.

\end{abstract}
\section{Introduction}
Pulmonary Hypertension (PH) is a severe disease characterised by elevated pressure in the Main Pulmonary Artery. The gold standard for detecting PH is Right Heart Catheterization (RHC), an invasive procedure enabling the assessment of pulmonary circulation characteristics including mean Pulmonary Artery Pressure (mPAP), cardiac output, or stroke volume. PH is diagnosed when mPAP exceeds 20 mmHg \cite{MHoeper2017}. Regrettably, as an invasive procedure, RHC is prone to complications as well as it requires specialized equipment, trained staff and patient preparation. As a consequence, researchers have adopted Machine Learning (ML) models to diagnose PH from demographics data \cite{Leha2019}, measurements from cardiac computed tomography \cite{Huang2020}, physics-based models parameters \cite{Lungu2016} and Magnetic Resonance Imaging-derived (MRI) measurements \cite{alabed2022validation,grzeszczyk2022noninvasive}. In this study, we take a new route and we explore the potential of estimating mPAP directly from Cardiac MRI (CMR) videos of one cardiac cycle collected at two planes: 4 Chamber (4CH) and Short-Axis (SA) using Deep Learning.

In recent years, 3D Convolutional Neural Networks (CNNs) \cite{carreira2017quo,tran2018closer} and Vision Transformers \cite{Arnab_2021_ICCV,liu2022video} were the go-to standard for video processing. More recently the emergence of Multi-layer Perceptrons (MLP)-based models \cite{chen2022cyclemlp,tolstikhin2021mlp,touvron2022resmlp} has led to the development of their 3D counterparts \cite{qiu2022mlp}. In the medical domain, the common approach of collecting imaging and tabular data resulted in the introduction of fusion methods. Traditional methods involve concatenating tabular features in the network's final layers \cite{holste2021end}, but this restricts the knowledge transfer with imaging data. To enhance the interaction between imaging and tabular data Duanmu \textit{et al.} \cite{duanmu2020prediction} proposed a method in which imaging features are channel-wise multiplied with tabular data at different stages of CNN. P\"{o}lsterl \textit{et al.} \cite{polsterl2021combining} introduced Dynamic Affine Feature Map Transform (DAFT) module, an extension of the Feature-wise Linear Modulation (FiLM) \cite{perez2018film} layer for scaling and shifting feature maps based on tabular data. Grzeszczyk \textit{et al.} \cite{grzeszczyk2023tabattention} proposed TabAttention, a module for learning channel, spatial and temporal attention conditioned on tabular data. In this work, we enhance the predictive capabilities of imaging models by introducing tabular data in the form of demographic features and measurements from MRI (e.g. right ventricular ejection fraction).

Traditionally, MLPs were used for tabular data processing and more recently for image analysis, however, no approach combines both modalities. Inspired by all-MLP architectures, we present TabMixer, a novel module facilitating the mixing of tabular data with imaging features. TabMixer enriches the video processing backbones by enabling interaction between tabular embeddings and imaging features across spatial, temporal and channel dimensions. Unlike convolutional or self-attention-based methods, TabMixer relies solely on MLPs, allowing the fusion of information stored in tabular data with imaging features during spatial, temporal and channel mixing. We evaluate TabMixer integrated with different backbones (CNNs, Transformer, all-MLP) on the mPAP prediction task and show its competitiveness with existing methods. Our contributions are threefold: (1) we introduce TabMixer, a first module akin to MLP designed for the integration of imaging feature maps with tabular data, (2) we provide a comprehensive evaluation and comparison of methods that utilize imaging and/or tabular data for predicting mPAP, and (3) to our knowledge, this is the first demonstration of noninvasive mPAP estimation directly from CMR videos. In addition, we present a comparative analysis of the 4 Chamber (4CH) and Short-Axis (SA) planes for this estimation.

\begin{figure}[t!]
    \centering
    \includegraphics[width=\textwidth]{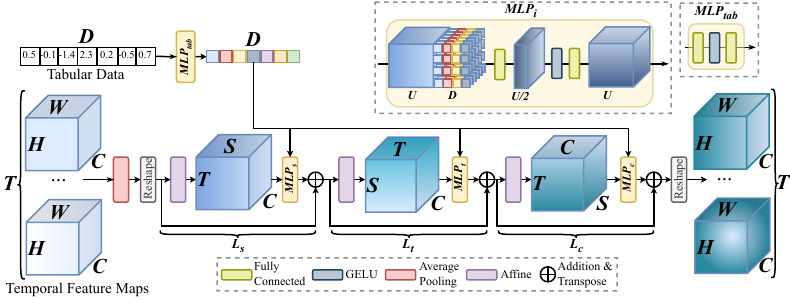}
    \caption{The overview of the proposed TabMixer. TabMixer applies mixing operations over spatial, temporal and channel dimensions. Leveraging MLPs, TabMixer can interchange information between imaging and tabular features.}
    \label{fig:overview}
\end{figure}

\section{Method}

In this section, we introduce the building blocks of the TabMixer module. We present the rationale behind employing MLP architecture for imaging and tabular data. Finally, we describe the integration of TabMixer with vision backbones.

The overview of TabMixer is presented in Fig. \ref{fig:overview}. Let $X_{0} \in \R^{1 \times T_{0} \times H_{0} \times W_{0}}$ denote the grayscale CMR video comprising $T_{0}$ frames with $H_{0}$ height and $W_{0}$ width. The vision backbone generates intermediate temporal feature maps $X \in \R^{C \times T \times H \times W}$ where $C$ is the number of channels. TabMixer refines these temporal feature maps through spatial, temporal and channel mixing with the aid of tabular data $Tab \in \R^{D}$. We follow the architectural principles of all-MLP and Transformer networks in the design of this module \cite{yu2022metaformer}. 

Initially, to reduce dimensionality and the number of parameters, the intermediate temporal feature maps $X$ are embedded via average pooling and reshaping operations into a features cube $X' \in \R^{C \times T \times S}$ where $S$ is $\frac{(H \times W)}{4}$. This structure of $X'$  enables the interchange of spatial, temporal and channel information with tabular data via three sub-layers $L_{s}$, $L_{t}$ and $L_{c}$, sequentially operating across every dimension. Each $L_{i}$ component comprises an Affine transformation, $MLP_{i}$ layer and a skip-connection followed by permutation operation to facilitate processing over the next dimension. Finally, the processed features are reshaped and linearly upsampled back to the input shape, thereby returning refined temporal feature maps integrated with tabular information.

\noindent \textbf{Mixing with Multi-Layer Perceptron.} In many networks, a normalization layer precedes model-specific layers (e.g. attention) to stabilize training. Inspired by \cite{touvron2022resmlp}, instead of Layer Normalization, we perform Affine transformation in $L_{i}$, to rescale and shift inputs element-wise without depending on batch statistics:
\begin{equation}
       Aff_{\alpha,\beta}(X') = Diag(\alpha)X'+\beta
\end{equation}

\noindent After the affine transformation, the three-dimensional cube is processed with $MLP_i$ block consisting of two MLP layers and GELU \cite{hendrycks2016gaussian} activation:

\begin{equation}
       Y = MLP_{i,2}(GELU(MLP_{i,1}(Z)))
\end{equation}

Here $Z$ is the input to the $MLP_{i}$ and $Y$ is the output of the same size. The $MLP_{i}$ block enables mixing information across the input's last dimension and is shared across all vectors in this dimension as in \cite{tolstikhin2021mlp}. The first MLP compresses the last dimension by a factor of two to create a bottleneck, while the second MLP expands it back to the initial size (as shown in Fig. \ref{fig:overview}). Subsequently, we employ the skip connection with the $L_{i}$ input. As $MLP_{i+1}$ operates on different dimension, the final operation within $L_{i}$ entails axis permutation. The overall dimensionality evolves in TabMixer as follows:

\begin{equation}
       \stackrel{\text{input}}{\R^{C \times T \times H \times W}} \rightarrow  \stackrel{L_{s}}{\R^{C \times T \times S}}  \rightarrow \stackrel{L_{t}}{\R^{C \times S \times T}} \rightarrow \stackrel{L_{c}}{\R^{S \times T \times C}} \rightarrow  \stackrel{\text{output}}{\R^{C \times T \times H \times W}}
\end{equation}

\noindent \textbf{Tabular data within TabMixer.} To incorporate tabular information within spatial, temporal and channel processing we embed tabular data with $MLP_{tab}$ ($Tab \in \R^{D} \rightarrow Tab' \in \R^{D}$) duplicating the structure of $MLP_{i}$ blocks. Inspired by the use of MLPs in tabular Deep Learning models and all-MLP networks for vision, we concatenate tabular embedding with the $X'$ features cube before processing it via $MLP_{i}$. This concatenation involves repeating tabular embedding over the first and second axes of $X'$, augmenting the input to $MLP_{i}$ by $D$ values from tabular embedding (e.g. $\R^{C \times T \times S+D}$ for $MLP_{s}$). This approach enables the fusion of tabular features with imaging ones across all dimensions, as shown in Fig. \ref{fig:overview}. The processing of feature maps with tabular data can be expressed as:

\begin{align}
       &TabMixer(X, Tab) = U(L_{c}(L_{t}(L_{s}(R(X), Tab'), Tab'), Tab'))\\
       &Tab' = MLP_{tab}(Tab)
\end{align}

\noindent Here, $R$ and $U$ denote the reshaping with pooling and upsampling for simplicity. The design of TabMixer enables its integration with networks based on a hierarchical structure (e.g. ResNet-18, 3D-MLP) yielding intermediate feature maps.

\begin{figure}[t!]
  \includegraphics[width=\linewidth]{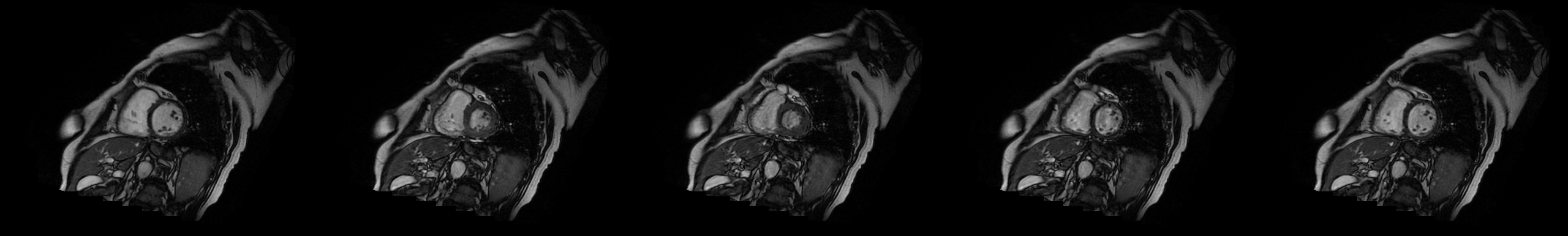}
  \includegraphics[width=\linewidth]{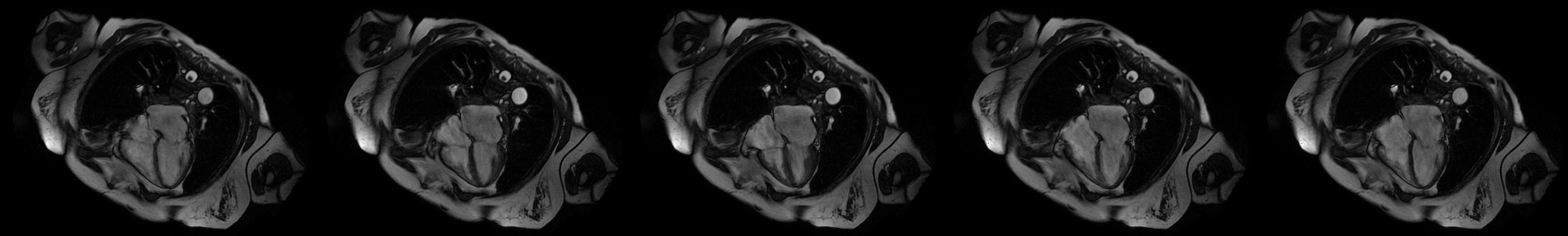}
\caption{Short-Axis (top) and corresponding 4 Chamber (bottom) frames from CMR depicting a cardiac cycle from left to right.}
\label{fig:frames}
\end{figure}

\section{Experiments and results}
This section describes the PH dataset used for mPAP estimation and outlines the implementation details of TabMixer. We benchmark our method against various state-of-the-art modules and compare its performance with imaging-only and tabular-only ML and Deep Learning models. Additionally, we conduct an ablation study and demonstrate the importance of TabMixer's main components.

\noindent \textbf{Dataset.} This study was approved by the Ethics Committee. The dataset employed comprises 1821 CMR studies (3642 videos of SA and 4CH planes) matched with invasively measured mPAP and is part of the ASPIRE Registry (Assessing the Severity of Pulmonary Hypertension In a Pulmonary Hypertension REferral Centre) \cite{Hurdman2012}. It contains data from 1708 patients (1045 females and 663 males, 64 $\pm$ 14 years old) suspected of PH who underwent both MRI and RHC within 48 hours (some patients had repeated the procedures over the years). To ensure a fair comparison across planes and methods based on tabular data we extract data samples from the ASPIRE Registry containing  SA and 4CH CMR videos, and select tabular features (demographics and measurements from MRI) with no more than 200 empty values. We remove data samples without tabular features. The MRI videos, capturing one cardiac cycle (up to 50 frames) were acquired during a breath hold in the supine position, utilizing devices from various vendors, including GE, Philips and Siemens. Exemplary frames from corresponding SA and 4CH planes are shown in Fig. \ref{fig:frames}.

\bgroup
\def\arraystretch{1.01}%
\begin{table}[ht!]
    \caption{The performance of imaging (I.) and/or tabular (T.) methods. We \textbf{highlight} best and \underline{underline} second MAE [mmHg], RMSE [mmHg], MAPE [\%]. We indicate the statistically significant difference (two-tailed paired t-test p-value $<0.01$) with TabMixer over the baseline on both 4CH and SA planes with *.}
    \label{tab:resultssota}
    \begin{center}
    \begin{tabular}{l|c|c|c|c|c||c|c|c}
       
       \textbf{Method} &\textbf{I.} & \textbf{T.}& \textbf{MAE} & \textbf{RMSE} & \textbf{MAPE} & \textbf{MAE} & \textbf{RMSE} & \textbf{MAPE}\\ \hline
       LR \cite{scikit-learn}                                             & \xmark & \cmark & $8.50$ & $11.01$ & $23.35$ & $-$ & $-$ & $-$\\ 
       MLP\cite{scikit-learn}                                             & \xmark & \cmark & $8.25$ & $10.87$ & $22.51$ & $-$ & $-$ & $-$\\
       Trompt \cite{chen2023trompt}                                       & \xmark & \cmark & $8.12$ & $10.61$ & $22.38$ & $-$ & $-$ & $-$\\
       XGBoost \cite{xgboost}                                             & \xmark & \cmark & $7.82$ & $10.11$ & $21.84$ & $-$ & $-$ & $-$\\ 
       ResNet$_{tab}$ \cite{gorishniy2021revisiting}                      & \xmark & \cmark & $7.65$ & $9.96$ & $21.10$ & $-$ & $-$ & $-$\\ 
       FT-Transformer \cite{gorishniy2021revisiting}                      & \xmark & \cmark & $7.62$ & $10.03$ & $\underline{20.89}$ & $-$ & $-$ & $-$\\
       GBDT \cite{scikit-learn}                                           & \xmark & \cmark & $\underline{7.58}$ & $\underline{9.78}$ & $\textbf{20.85}$ & $-$ & $-$ & $-$\\ 
       RF \cite{scikit-learn}                                             & \xmark & \cmark & $\textbf{7.47}$ & $\textbf{9.61}$ & $20.97$ & $-$ & $-$ & $-$\\ \hline \hline
       
       \multicolumn{3}{c|}{} & \multicolumn{3}{c||}{\textbf{Short-Axis}}    & \multicolumn{3}{c}{\textbf{4 Chamber}} \\ \hline
             
       Video Swin \cite{liu2022video}                               & \cmark & \xmark & $11.65$ & 14.10 & $34.82$ & $12.01$ & 14.47 & $36.21$\\ 
       $\>$ + Concat                                     & \cmark & \cmark & $10.09$ & 12.34 & $29.99$ & $10.17$ & 12.43 & $30.53$\\
       $\>$ + FiLM \cite{perez2018film}                  & \cmark & \cmark & $\underline{7.98}$ & 10.42 & $\underline{22.62}$ & $8.10$ & \underline{10.43} & $\underline{22.80}$\\
       $\>$ + DAFT \cite{polsterl2021combining}          & \cmark & \cmark & $11.88$ & 14.41 & $36.64$ & $\underline{8.09}$ & 10.51 & $22.96$\\ 
       $\>$ + TabAttention \cite{grzeszczyk2023tabattention}   & \cmark & \cmark & $8.02$ & \underline{10.38} & $23.07$ & $8.60$ & 11.11 & $25.12$\\
       $\>$ + \textbf{TabMixer*}                          & \cmark & \cmark & $\textbf{7.86}$ & \textbf{10.22} & $\textbf{22.25}$ & $\textbf{7.92}$ & \textbf{10.34} & $\textbf{22.30}$\\ \hline

        MLP-3D \cite{qiu2022mlp}                                     & \cmark & \xmark & $11.84$ & 14.24 & $35.72$ & $12.04$ & 14.51 & $36.09$\\
       $\>$ + Concat                                     & \cmark & \cmark & $9.59$ & 11.97 & $27.87$ & $10.23$ & 12.45 & $30.13$\\
       $\>$ + FiLM \cite{perez2018film}                  & \cmark & \cmark & $\underline{8.13}$ & 10.67 & $\underline{22.62}$ & $\underline{8.01}$ & \underline{10.54} & $\textbf{22.28}$\\
       $\>$ + DAFT \cite{polsterl2021combining}          & \cmark & \cmark & $11.41$ & 13.83 & $33.48$ & $8.95$ & 11.54 & $24.51$\\  
       $\>$ + TabAttention \cite{grzeszczyk2023tabattention}   & \cmark & \cmark & $8.29$ & \underline{10.62} & $24.55$ & $8.48$ & 10.72 & $24.48$\\ 
       $\>$ + \textbf{TabMixer*}                          & \cmark & \cmark & $\textbf{7.80}$ & \textbf{10.20} & $\textbf{22.02}$ & $\textbf{7.97}$ & \textbf{10.32} & $\underline{22.76}$\\ \hline

       ResNet$_{DAFT}$ \cite{polsterl2021combining}                     & \cmark & \xmark & $12.09$ & 14.53 & $35.48$ & $12.24$ & 14.67 & $34.91$\\ 
       $\>$ + Concat                                         & \cmark & \cmark & $10.50$ & 12.82 & $31.97$ & $9.79$ & 11.81 & $28.93$\\
       $\>$ + FiLM \cite{perez2018film}                      & \cmark & \cmark & $\underline{8.24}$ & \underline{10.69} & $23.38$ & $8.07$ & 10.67 & $22.41$\\
       $\>$ + DAFT \cite{polsterl2021combining}              & \cmark & \cmark & $8.25$ & 10.98 & $\underline{22.06}$ & $\underline{7.87}$ & \underline{10.51} & $\underline{22.10}$\\ 
       $\>$ + TabAttention \cite{grzeszczyk2023tabattention} & \cmark & \cmark & $8.54$ & 11.19 & $23.65$ & $8.26$ & 10.83 & $23.29$\\ 
       $\>$ + \textbf{TabMixer*}                              & \cmark & \cmark & $\textbf{7.82}$ & \textbf{10.22} & $\textbf{22.04}$ & $\textbf{7.68}$ & \textbf{10.24} & $\textbf{21.50}$\\ \hline
       
       ResNet-18 \cite{tran2018closer}                              & \cmark & \xmark & $7.85$ & 10.21 & $21.71$ & $8.05$ & 10.47 & $23.34$\\ 
       $\>$ + Concat                                     & \cmark & \cmark & $\textbf{7.22}$ & \textbf{9.36} & $\textbf{19.92}$ & $8.05$ & 10.43 & $22.74$\\
       $\>$ + FiLM \cite{perez2018film}                  & \cmark & \cmark & $9.05$ & 11.63 & $24.58$ & $8.91$ & 11.44 & $24.40$\\
       $\>$ + DAFT \cite{polsterl2021combining}          & \cmark & \cmark & $8.93$ & 11.53 & $25.43$ & $8.14$ & 10.39 & $23.00$\\  
       $\>$ + TabAttention \cite{grzeszczyk2023tabattention}   & \cmark & \cmark & $7.92$ & 10.03 & $22.77$ & $\underline{7.72}$ & \underline{9.85} & $\underline{21.76}$\\
       $\>$ + \textbf{TabMixer}                          & \cmark & \cmark & $\underline{7.56}$ & \underline{9.94} & $\underline{21.57}$ & $\textbf{7.53}$ & \textbf{9.91} & $\textbf{21.75}$\\ \hline
       
       I3D \cite{carreira2017quo}                                   & \cmark & \xmark & $7.21$ & $9.39$ & $\underline{19.11}$ & $8.07$ & 10.29 & $23.05$\\ 
       $\>$ + Concat                                     & \cmark & \cmark & $7.56$ & $9.72$ & $20.76$ & $7.86$ & 9.83 & $20.86$\\
       $\>$ + FiLM \cite{perez2018film}                  & \cmark & \cmark & $8.53$ & $11.11$ & $22.48$ & $9.10$ & 11.66 & $24.34$\\
       $\>$ + DAFT \cite{polsterl2021combining}          & \cmark & \cmark & $\underline{7.18}$ & $\underline{9.25}$ & $19.21$ & $\underline{7.74}$ & 9.83 & $\underline{20.56}$\\
       $\>$ + TabAttention \cite{grzeszczyk2023tabattention}   & \cmark & \cmark & $7.34$ & $9.49$ & $21.34$ & $7.77$ & \underline{9.76} & $21.98$\\
       $\>$ + \textbf{TabMixer*}                          & \cmark & \cmark & $\textbf{6.66}$ & \textbf{8.64} & $\textbf{18.90}$ & \textbf{7.19} & \textbf{9.31} & \textbf{20.01}\\ 
            
    \end{tabular}
    \end{center}
\end{table}

\noindent \textbf{Implementation details.} We implement all Deep Learning models using PyTorch and train them on NVIDIA A100 80 GPU for 100 epochs. To minimize the Mean Squared Error loss of the mPAP prediction we use AdamW optimizer \cite{loshchilov2017decoupled} with a cosine annealing rate scheduler. An initial learning rate is set to $1 \times 10^{-4}$, L2 regularization to $1 \times 10^{-5}$ and batch size to 8. We split the dataset into training, validation and testing sets with 1299, 217 and 305 samples respectively, ensuring that each patient's data is present in only one set. We stratify the split based on the four mPAP bins ($\leq20$ mmHg, $\leq25$ mmHg, $\leq30$ mmHg, $>30$ mmHg). We resample the pixel spacing of all frames to 0.9375 mm $\times$ 0.9375 mm, pad them to 512x512 pixels (the maximum frame size in the dataset) and resize entire videos to 192x192 pixels with 16 frames. During training, we apply data augmentation methods, including translation, rotation, contrast adjustment, gaussian noise, intensity scaling and shifting. We one-hot categorical tabular features and standardize the numerical ones to a mean value of 0 and a standard deviation of 1. We retain tabular features exhibiting statistical significance measured with f-regression \cite{scikit-learn} and evaluate the performance in estimating mPAP using Mean Absolute Error [mmHg] (MAE), Root Mean Squared Error [mmHg] (RMSE) and Mean Absolute Percentage Error [\%] (MAPE). Due to space limitations, we present mean results here, while standard deviations and the description of 29 tabular features are in the supplementary material.

\noindent \textbf{Comparison with state-of-the-art methods.} To compare TabMixer with several methods for imaging and tabular data (Concatenation, FiLM \cite{perez2018film}, DAFT \cite{polsterl2021combining}, TabAttention \cite{grzeszczyk2023tabattention}) we use five vision backbones: 3D CNNs (ResNet-18 \cite{tran2018closer}, I3D \cite{carreira2017quo}, ResNet$_{DAFT}$ \cite{polsterl2021combining}), Transformer (Video Swin Transformer \cite{liu2022video}) and all-MLP network (MLP-3D \cite{qiu2022mlp}). In all backbones, we insert tabular modules before the global average pooling layer. We conduct the mPAP prediction on SA and 4CH planes to find the better one for this task. Additionally, we test the performance of mPAP estimation by methods based solely on tabular data. We experiment with Deep Learning models (Trompt \cite{chen2023trompt}, ResNet$_{tab}$ \cite{gorishniy2021revisiting}, FT-Transformer \cite{gorishniy2021revisiting}, MLP)  and ML - Linear Regression (LR), XGBoost \cite{xgboost}, Gradient Boosting Decision Trees (GBDT), Random Forest (RF). Results are presented in Table \ref{tab:resultssota}. RF is the best-performing tabular-only method with MAE of $7.47\pm6.05$, while I3D trained on the SA plane achieves the lowest error out of imaging-only methods ($7.21\pm6.01$). TabMixer outperforms all modules when tested on the 4CH plane and has the best results for four out of five vision backbones on the SA plane (the improvement over the four baselines on both planes is statistically significant with a p-value below 0.002). Notably, the combination of I3D with TabMixer and SA plane reaches the lowest error with MAE of $6.66\pm5.51$.

\noindent \textbf{Ablation study.} We conduct ablation experiments to validate the key components of TabMixer (Table \ref{tab:ablation}). We perform validation of our proposed method when integrated with the I3D model. The incorporation of spatial, temporal and channel mixing components provides the lowest prediction error. Conversely, removing any of those blocks or tabular data mixing increases the error while still retaining performance gain over the baseline I3D.

\bgroup
\def\arraystretch{1}%
\begin{table}[t!]
    \caption{The ablation study of the key components of the proposed TabMixer. The first row is the baseline I3D model. The next rows refer to I3D combined with TabMixer without tabular data or one of the mixing components and the last row is full TabMixer.}
    \label{tab:ablation}
    \begin{center}
    \begin{tabular}{l|c|c|c|c|c}
       
       \textbf{Method} &\textbf{I.} & \textbf{T.}& \textbf{MAE} & \textbf{RMSE} & \textbf{MAPE} \\ \hline

       I3D \cite{carreira2017quo}                                   & \cmark & \xmark & $7.21 \pm 6.01$ & $9.39$ & $19.11 \pm 19.62$\\ 
       $\>$ + TM w/o tabular data   & \cmark & \xmark & $7.17\pm5.71$ & 9.17 & $19.97\pm22.84$ \\
       $\>$ + TM w/o channel mixing   & \cmark & \cmark & $7.08\pm5.73$ & 9.11 & $19.80\pm22.72$ \\
       $\>$ + TM w/o spatial \& temporal mixing   & \cmark & \cmark & $7.01\pm6.01$ & 9.24 & $19.07\pm20.00$ \\
       $\>$ + TM w/o spatial mixing   & \cmark & \cmark & $7.00\pm5.78$ & 9.08 & $19.11\pm20.12$ \\
       $\>$ + TM w/o temporal mixing   & \cmark & \cmark & $6.89\pm5.83$ & 9.02 & $19.60\pm21.72$\\
       $\>$ + \textbf{TabMixer} (TM)       & \cmark & \cmark & $\textbf{6.66} \pm 5.51$ & \textbf{8.64} & $\textbf{18.90}\pm20.83$\\ 
            
    \end{tabular}
    \end{center}
\end{table}

\bgroup
\def\arraystretch{1.01}%
\begin{table}[t!]
    \caption{The parameters count and inference time of modules assuming feature maps shape from the I3D (1024, 4, 6, 6 - C, T, H, W) and the number of tabular features of 29. We also provide measures for TabMixer without channel mixing (TM w/o CM).}
    \label{tab:performance}
    \begin{center}
    \begin{tabular}{l|c|c|c|c|c}
       
       \textbf{Method} & \textbf{FiLM} & \textbf{DAFT} & \textbf{TabAttention} & \textbf{TabMixer} & \textbf{TM w/o CM} \\ \hline
       \textbf{Params [M]} & 0.015 & 0.022 & 0.202 & 1.070 & 0.001\\ \hline
       \textbf{Inference time} [ms] & 0.04 & 0.35 & 7.57 & 1.41 &  0.97

    \end{tabular}
    \end{center}
\end{table}

\begin{figure}[t!]
    \centering
    \includegraphics[width=\textwidth]{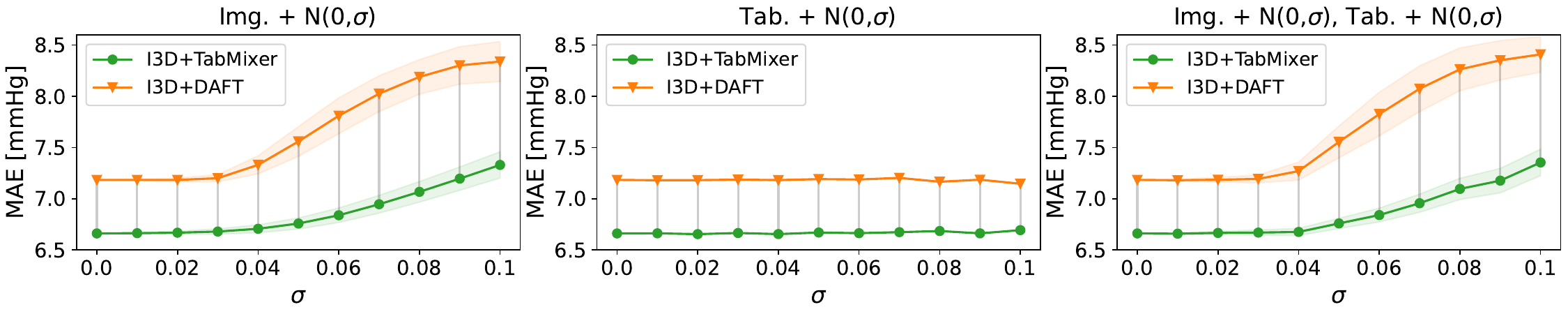}
    \caption{The comparison of resistance to noisy imaging (left), tabular (middle) data and both (right). We compare the performance of I3D+TabMixer and the second best-performing method I3D+DAFT.}
    \label{fig:noise}
\end{figure}

\section{Discussion and Conclusions}
To the best of our knowledge, we are the first to directly estimate mPAP from CMR videos. Notably, our analysis indicates the superiority of the SA plane over 4CH in predicting mPAP which is the same CMR plane used for estimation of ejection fraction or stroke volume. Inspired by prior works combining imaging and tabular data, as well as all-MLP networks, we introduced TabMixer - a module for combining CMR and tabular data in 3D networks. TabMixer enhances the predictive capabilities of all investigated networks. It worked best with the I3D model (the improvement over the baseline is statistically significant with p-value$\approx$0.002). However, while other methods exhibited improvement, they did not surpass the RF method. RF outperformed Deep Learning methods for tabular data like Trompt or FT-Transformer which is consistent with the challenges Deep Learning models face in effectively leveraging tabular data \cite{shwartz2022tabular}. This highlights the importance of the right combination of the vision backbone and tabular module to the task.  


To better understand the impact of imaging and tabular data, we experiment with noisy data applied to two best-performing methods I3D+TabMixer and I3D+DAFT. As shown in Fig. \ref{fig:noise}, both methods exhibit resilience to noise in tabular data with I3D+TabMixer showcasing greater robustness to noisy videos owing to its superior representation learning facilitated by tabular and imaging data mixing.

Our method exhibits certain limitations. Firstly, it was not tested on the external dataset and to prove its generalizability, further study is needed. Secondly, we experimented with estimations based on single CMR planes. The combination of both 4CH and SA videos could yield an additional improvement in the quality of predictions. Finally, TabMixer is not a lightweight module, due to the extensive use of MLP layers. As the spatial, temporal, and channel dimensions increase, the number of trainable parameters within the MLP layers escalates quadratically. For example, for I3D and 1024 channels in the final temporal feature maps the number of trainable parameters of TabMixer is higher than those of FiLM or DAFT as shown in Table \ref{tab:performance}. However, TabMixer without channel mixing has much less trainable parameters but also outperforms all other modules in terms of MAE and RMSE as presented in the ablation study (Table \ref{tab:ablation}).

In summary, this paper addresses the challenge of noninvasive mPAP estimation. We have introduced the TabMixer module, which facilitates the integration of tabular and imaging data. To our knowledge, this represents the first application of MLP models to merge these two data modalities. Our approach holds considerable promise for clinical workflows, particularly in mPAP estimation. By offering a noninvasive alternative, it has the potential to transform the field, conceivably obviating the need for high-risk invasive RHC procedures.

\subsubsection{\ackname}
This research was funded in whole or in part by National Science Centre, Poland 2023/49/N/ST6/01841. For the purpose of Open Access, the author has applied a CC-BY public copyright licence to any Author Accepted Manuscript (AAM) version arising from this submission. This work is supported by the EU's Horizon 2020 programme (grant no. 857533, Sano) and the Foundation for Polish Science's International Research Agendas programme, co-financed by the EU under the European Regional Development Fund and the Polish Ministry of Science and Higher Education (contract no. MEiN/2023/DIR/3796).

\subsubsection{\discintname}
The authors have no conflicts of interest to declare.
\bibliographystyle{splncs04}
\bibliography{Paper-2329}

\begin{thebibliography}{10}
\providecommand{\url}[1]{\texttt{#1}}
\providecommand{\urlprefix}{URL }
\providecommand{\doi}[1]{https://doi.org/#1}

\bibitem{alabed2022validation}
Alabed, S., Alandejani, F., Dwivedi, K., Karunasaagarar, K., Sharkey, M., Garg, P., de~Koning, P.J., T{\'o}th, A., Shahin, Y., Johns, C., et~al.: Validation of artificial intelligence cardiac mri measurements: relationship to heart catheterization and mortality prediction. Radiology  \textbf{305}(1),  68--79 (2022)

\bibitem{Arnab_2021_ICCV}
Arnab, A., Dehghani, M., Heigold, G., Sun, C., Lu\v{c}i\'c, M., Schmid, C.: Vivit: A video vision transformer. In: Proceedings of the IEEE/CVF International Conference on Computer Vision (ICCV). pp. 6836--6846 (October 2021)

\bibitem{carreira2017quo}
Carreira, J., Zisserman, A.: Quo vadis, action recognition? a new model and the kinetics dataset. In: proceedings of the IEEE Conference on Computer Vision and Pattern Recognition. pp. 6299--6308 (2017)

\bibitem{chen2023trompt}
Chen, K.Y., Chiang, P.H., Chou, H.R., Chen, T.W., Chang, T.H.: Trompt: Towards a better deep neural network for tabular data. arXiv preprint arXiv:2305.18446  (2023)

\bibitem{chen2022cyclemlp}
Chen, S., Xie, E., GE, C., Chen, R., Liang, D., Luo, P.: Cycle{MLP}: A {MLP}-like architecture for dense prediction. In: International Conference on Learning Representations (2022)

\bibitem{xgboost}
Chen, T., Guestrin, C.: {XGBoost}: A scalable tree boosting system. In: Proceedings of the 22nd ACM SIGKDD International Conference on Knowledge Discovery and Data Mining. pp. 785--794. KDD '16, ACM, New York, NY, USA (2016). \doi{10.1145/2939672.2939785}

\bibitem{duanmu2020prediction}
Duanmu, H., Huang, P.B., Brahmavar, S., Lin, S., Ren, T., Kong, J., Wang, F., Duong, T.Q.: Prediction of pathological complete response to neoadjuvant chemotherapy in breast cancer using deep learning with integrative imaging, molecular and demographic data. In: Medical Image Computing and Computer Assisted Intervention--MICCAI 2020: 23rd International Conference, Lima, Peru, October 4--8, 2020, Proceedings, Part II 23. pp. 242--252. Springer (2020)

\bibitem{gorishniy2021revisiting}
Gorishniy, Y., Rubachev, I., Khrulkov, V., Babenko, A.: Revisiting deep learning models for tabular data. Advances in Neural Information Processing Systems  \textbf{34},  18932--18943 (2021)

\bibitem{grzeszczyk2022noninvasive}
Grzeszczyk, M.K., Sat{\l}awa, T., Lungu, A., Swift, A., Narracott, A., Hose, R., Trzcinski, T., Sitek, A.: Noninvasive estimation of mean pulmonary artery pressure using mri, computer models, and machine learning. In: International Conference on Computational Science. pp. 14--27. Springer (2022)

\bibitem{grzeszczyk2023tabattention}
Grzeszczyk, M.K., et~al.: Tabattention: Learning attention conditionally on tabular data. In: International Conference on Medical Image Computing and Computer-Assisted Intervention. pp. 347--357. Springer (2023)

\bibitem{hendrycks2016gaussian}
Hendrycks, D., Gimpel, K.: Gaussian error linear units (gelus). arXiv preprint arXiv:1606.08415  (2016)

\bibitem{MHoeper2017}
Hoeper, M.M., et~al.: Pulmonary hypertension. Dtsch Arztebl Int  \textbf{114},  73--84 (2017). \doi{10.3238/arztebl.2017.0073}

\bibitem{holste2021end}
Holste, G., Partridge, S.C., Rahbar, H., Biswas, D., Lee, C.I., Alessio, A.M.: End-to-end learning of fused image and non-image features for improved breast cancer classification from mri. In: Proceedings of the IEEE/CVF International Conference on Computer Vision. pp. 3294--3303 (2021)

\bibitem{Huang2020}
Huang, L., Li, J., Huang, M., Zhuang, J., Yuan, H., Jia, Q., Zeng, D., Que, L., Xi, Y., Lin, J., Dong, Y.: Prediction of pulmonary pressure after glenn shunts by computed tomography-based machine learning models. European Radiology  \textbf{30},  1369--1377 (2020). \doi{10.1007/s00330-019-06502-3}

\bibitem{Hurdman2012}
Hurdman, J., Condliffe, R., Elliot, C., Davies, C., Hill, C., et~al.: Aspire registry: Assessing the spectrum of pulmonary hypertension identified at a referral centre. European Respiratory Journal  \textbf{39},  945--955 (4 2012). \doi{10.1183/09031936.00078411}

\bibitem{Leha2019}
Leha, A., Hellenkamp, K., Unsöld, B., Mushemi-Blake, S., Shah, A.M., Hasenfuß, G., Seidler, T.: A machine learning approach for the prediction of pulmonary hypertension. PLoS ONE  \textbf{14} (10 2019). \doi{10.1371/journal.pone.0224453}

\bibitem{liu2022video}
Liu, Z., Ning, J., Cao, Y., Wei, Y., Zhang, Z., Lin, S., Hu, H.: Video swin transformer. In: Proceedings of the IEEE/CVF conference on computer vision and pattern recognition. pp. 3202--3211 (2022)

\bibitem{loshchilov2017decoupled}
Loshchilov, I., Hutter, F.: Decoupled weight decay regularization. arXiv preprint arXiv:1711.05101  (2017)

\bibitem{Lungu2016}
Lungu, A., Swift, A.J., Capener, D., Kiely, D., Hose, R., Wild, J.M.: Diagnosis of pulmonary hypertension from magnetic resonance imaging-based computational models and decision tree analysis. Pulmonary Circulation  \textbf{6},  181--190 (6 2016). \doi{10.1086/686020}

\bibitem{scikit-learn}
Pedregosa, F., Varoquaux, G., Gramfort, A., Michel, V., Thirion, B., Grisel, O., Blondel, M., Prettenhofer, P., Weiss, R., Dubourg, V., Vanderplas, J., Passos, A., Cournapeau, D., Brucher, M., Perrot, M., Duchesnay, E.: Scikit-learn: Machine learning in {P}ython. Journal of Machine Learning Research  \textbf{12},  2825--2830 (2011)

\bibitem{perez2018film}
Perez, E., Strub, F., De~Vries, H., Dumoulin, V., Courville, A.: Film: Visual reasoning with a general conditioning layer. In: Proceedings of the AAAI conference on artificial intelligence. vol.~32 (2018)

\bibitem{polsterl2021combining}
P{\"o}lsterl, S., Wolf, T.N., Wachinger, C.: Combining 3d image and tabular data via the dynamic affine feature map transform. In: Medical Image Computing and Computer Assisted Intervention--MICCAI 2021: 24th International Conference, Strasbourg, France, September 27--October 1, 2021, Proceedings, Part V 24. pp. 688--698. Springer (2021)

\bibitem{qiu2022mlp}
Qiu, Z., Yao, T., Ngo, C.W., Mei, T.: Mlp-3d: A mlp-like 3d architecture with grouped time mixing. In: Proceedings of the IEEE/CVF Conference on Computer Vision and Pattern Recognition. pp. 3062--3072 (2022)

\bibitem{shwartz2022tabular}
Shwartz-Ziv, R., Armon, A.: Tabular data: Deep learning is not all you need. Information Fusion  \textbf{81},  84--90 (2022)

\bibitem{tolstikhin2021mlp}
Tolstikhin, I.O., et~al.: Mlp-mixer: An all-mlp architecture for vision. Advances in neural information processing systems  \textbf{34},  24261--24272 (2021)

\bibitem{touvron2022resmlp}
Touvron, H., et~al.: Resmlp: Feedforward networks for image classification with data-efficient training. IEEE Transactions on Pattern Analysis and Machine Intelligence  \textbf{45}(4),  5314--5321 (2022)

\bibitem{tran2018closer}
Tran, D., Wang, H., Torresani, L., Ray, J., LeCun, Y., Paluri, M.: A closer look at spatiotemporal convolutions for action recognition. In: Proceedings of the IEEE conference on Computer Vision and Pattern Recognition. pp. 6450--6459 (2018)

\bibitem{yu2022metaformer}
Yu, W., Luo, M., Zhou, P., Si, C., Zhou, Y., Wang, X., Feng, J., Yan, S.: Metaformer is actually what you need for vision. In: Proceedings of the IEEE/CVF conference on computer vision and pattern recognition. pp. 10819--10829 (2022)

\end{thebibliography}

\end{document}